\documentclass[preprint,12pt]{elsarticle}
\usepackage{graphicx}
\usepackage{amssymb}

\journal{Journal of Alloys and Compounds}

\begin{document}

\begin{frontmatter}

\title{The effect of Al doping on the structure and magnetism in cobaltite CaBaCo$_4$O$_7$}

\author[HMFL]{Youming Zou}
\author[HMFL]{Zhe Qu\corref{cor1}}\ead{zhequ@hmfl.ac.cn}
\author[HMFL]{Lei Zhang}
\author[HMFL]{Wei Ning}
\author[HMFL]{Langsheng Ling}
\author[USTC,HMFL]{Li Pi}
\author[HMFL,USTC]{Yuheng Zhang}
\address[HMFL]{High Magnetic Field Laboratory, Chinese Academy of Sciences, \\ Hefei, Anhui, 230031, China}
\address[USTC]{Hefei National Laboratory for Physical Sciences at the Microscale, \\ University of Science and Technology of China, Hefei, Anhui, 230026, China}
\cortext[cor1]{Corresponding author. Tel: +86-551-6559-5640; Fax: +86-551-6559-1149.}

\begin{abstract}
We report the effects of Al-doping on the structure and magnetic properties in CaBa(Co$_{1-x}$Al$_{x}$)$_4$O$_7$ (0$\leq$x$\leq$0.25). The system exhibits a structural transition from an orthorhombic symmetry to a hexagonal symmetry when the Al content exceeds $x =$ 0.1. The Curie temperature and the value of the magnetization decrease with increasing Al doping level, indicating that the ferrimagnetic ground state is gradually suppressed. The ground state eventually transits into a spin-glass state for $x >$ 0.1. Moreover, the short-range magnetic correlations, which occur at high temperatures in CaBaCo$_4$O$_7$, are found to be gradually suppressed with increasing Al content and eventually disappear for $x =$ 0.25. By comparing our results with other Co-site doping cases, we suggest that the lattice and the spin degrees of freedom are relatively decoupled in CaBaCo$_4$O$_7$.
\end{abstract}


\end{frontmatter}

\section{Introduction}


In recent years, there has been an increasing interest in geometrical frustrated magnets because of their exotic magnetic states such as spin liquids, spin ices and spin-glasses. \cite{GFM1,GFM2,GFM3,GFM4,GFM5,114book} In these materials, geometrical frustration results in highly degenerate or nearly degenerate ground states, making the system highly susceptible to external perturbations.


The recently discovered "114" cobaltites form a new class of geometrically frustrated magnets.\cite{114book,114A,114B,114C,114D,114E,114F} Their structure can be viewed as an alternative stacking of triangular and kagome layers build up by CoO$_4$ tetrahedra. Since the strong antiferromagnetic (AFM) Co-Co interactions are mediated by Co-O-Co superexchange pathways both in triangular and in kagome layers, they are expected to exhibit complex magnetic properties. For example, YBaCo$_4$O$_7$ was reported to undergo various magnetic transitions, such as spin-glass (SG) transition around $T_f \sim$ 66 K, \cite{114B} long-range AFM order below $T_N$ = 110 K, \cite{114G} and a magnetic transition with short-range correlations \cite{114H,114I}


Here we focus on the CaBaCo$_4$O$_7$, which has the largest orthorhombic distortion in the series. \cite{CaBaCo4O7SSC} Charge ordering is observed in this material, with Co$^{2+}$ sitting on two sites (Co2, Co3 sites) and Co$^{3+}$ sitting on the other two sites (Co1, Co4 sites). \cite{CaBaCo4O7neutron,CaBaCo4O7DFT} The system does not retain a disordered ground state since the geometric frustration is partially lifted by the large structural distortion and the charge ordering. It shows short-range magnetic correlations below $\sim$ 360 K and eventually enters a ferrimagnetic (FIM) ground state below $T_C \sim$ 60 K through a first-order transition, \cite{CaBaCo4O7SSC,CaBaCo4O7SRC} whose magnetic structure is found to consist of ferromagnetic (FM) zig-zag Co$^{2+}$ chains along $b$-axis that are antiferromagnetically coupled with Co$^{3+}$ cations. \cite{CaBaCo4O7neutron} In the FIM ground state, a spin-assisted ferroelectric state has been uncovered. \cite{CaBaCo4O7multiferro}



Chemical doping on cobalt sites could significantly tune the magnetic property of the CaBaCo$_4$O$_7$. Less than 3\% Zn impurities are found to induce a spectacular switching of the ground state from the FIM state to an AFM state with $T_N \sim$ 80 K. \cite{CaBaCo4O7Zn}
Such a spectacular switching of the ground state is attributed to the ordered doping of Zn$^{2+}$ at Co$^{2+}$ sites in the FM zig-zag Co$^{2+}$ chains, which could possibly induce the 180$^{\circ}$ flip of the spin on the neighbor Co$^{2+}$ cations and thus create an AFM state. \cite{CaBaCo4O7Zn}

In this work, we investigate the effect of replacing Co$^{3+}$ ions in CaBaCo$_4$O$_7$ by measuring the magnetic properties of CaBa(Co$_{1-x}$Al$_{x}$)$_4$O$_7$ with 0 $\leq x \leq$ 0.25. It is found that the system exhibits a structural transition from the orthorhombic symmetry to the hexagonal symmetry when the Al content exceeds $x =$ 0.1. The Curie temperature and the value of the magnetization decrease with increasing Al doping level, suggesting that the FIM ground state is gradually suppressed. The system eventually enters a spin-glass (SG) state for $x \geq$ 0.1. Moreover, the short-range magnetic correlations, which occur at high temperatures in CaBaCo$_4$O$_7$, are found to be gradually suppressed with increasing Al content and eventually disappear for $x =$ 0.25. By comparing our results with other doping cases, we suggest that the lattice and the spin degrees of freedom are relatively decoupled in CaBaCo$_4$O$_7$.

\section{Experiment}

Polycrystalline samples of CaBa(Co$_{1-x}$Al$_x$)$_4$O$_7$ with 0 $\leq$ x $\leq$ 0.25 were prepared by using the conventional solid-state reaction method described in Refs. \cite{CaBaCo4O7SSC,CaBaCo4O7SRC}. Stoichiometric proportions of high purity CaCO$_3$, BaCO$_3$, Al$_2$O$_3$ and Co$_3$O$_4$ were mixed and heated at 900 $^{o}C$ in air to decarbonation. They are then pelletized, and then sintered at 1100 $^{o}C$ in air for 12 hours and quenched to room temperature. The structure and the phase purity of the samples were checked by powder X-ray diffraction (XRD) at room temperature. Magnetization measurements were performed with a commercial superconducting quantum interference device (SQUID) magnetometer (Quantum Design MPMS 7T-XL) and a Physical Property Measurement System (Quantum Design PPMS-16T) equipped with a vibrating sample magnetometer (VSM). Since these cobaltites are sensitive to the oxygen content, we have carried out iodometric titration of our samples to measure the oxygen content. The results confirm that the oxygen stoichiometry is fixed to "O$_7$" within the limit of accuracy of $\pm$0.05. We also perform the EDX analysis on our samples and found that the cationic ratio is in agreement with the nominal composition.

\section{Results and Discussion}

\begin{figure}[b]
\center{\includegraphics[angle=-90,scale=1]{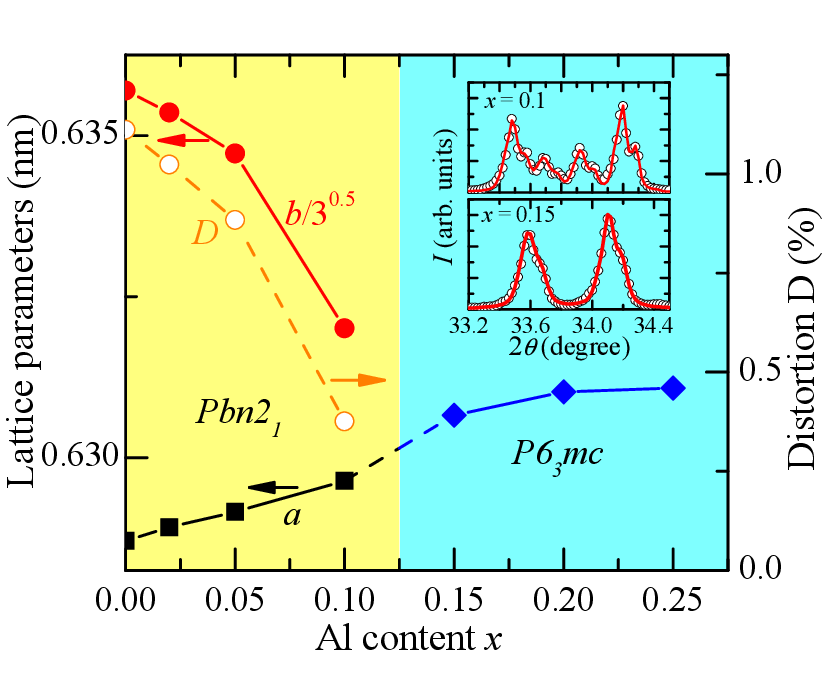}}
\caption{(Color online) The lattice parameters $a$, $b$ and the orthorhombic distortion $D = (b/\sqrt{3}-a)/a$ as function of the Al content $x$ for CaBa(Co$_{1-x}$Al$_{x}$)$_4$O$_7$. Insets show the enlarged XRD patterns and fitting results for $x =$ 0.1 and 0.15.}\label{fig:XRD}
\end{figure}

Figure \ref{fig:XRD} displays the evolution of the lattice parameters $a$ and $b$ for CaBa(Co$_{1-x}$Al$_x$)$_4$O$_7$ with 0 $\leq$ x $\leq$ 0.25. They are determined by performing the Rietveld refinements using a GSAS program. \cite{Rietveld,GSAS,EXPGUI} The parent compound CaBaCo$_4$O$_7$ has an orthorhombic structure ($Pbn2_1$ space group). \cite{CaBaCo4O7SSC} Upon Al doping, the orthorhombic distortion, which can be quantified as $D = (b/\sqrt{3}-a)/a$, \cite{CaBaCo4O7neutron} is significantly reduced (see Fig. \ref{fig:XRD}). For $x =$ 0.1, the heaviest doped sample retaining the orthorhombic symmetry, the value of $D$ is only half of that for the parent compound. When the doping level exceeds 10\%, the system shows a structure transition into a hexagonal symmetry ($P6_3mc$ space group). As a result, the splitting of the Bragg peaks between 2$\theta \sim$ 33$^\circ$ - 34$^\circ$, which is still evident for $x =$ 0.1, disappears for $x =$ 0.15.

\begin{figure}[b]
\center{\includegraphics[angle=0,scale=1]{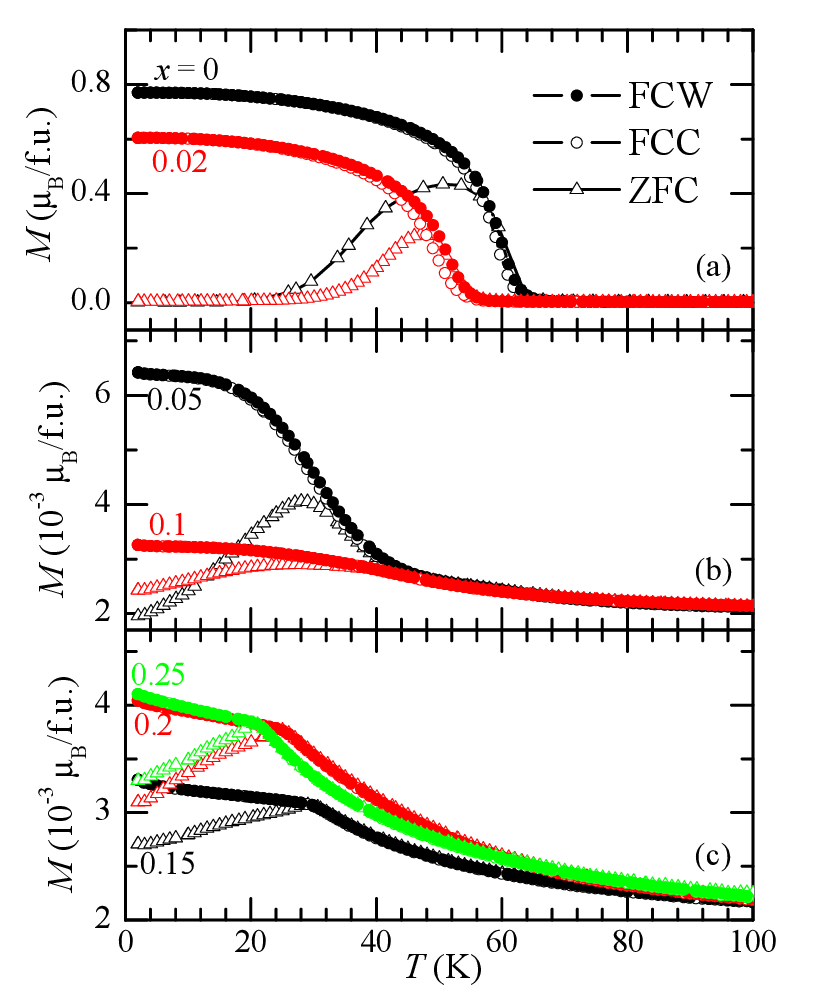}}
\caption{(Color online)The magnetization versus the temperature under 1000 Oe for CaBa(Co$_{1-x}$Al$_x$)$_4$O$_7$ with 0 $\leq x \leq$ 0.25 measured during warming after zero-field cooling sequence (ZFC), during field cooling sequence (FCC), and during warming after field cooling sequence (FCW).}\label{fig:MT}
\end{figure}

\begin{figure}[b]
\center{\includegraphics[angle=0,scale=1]{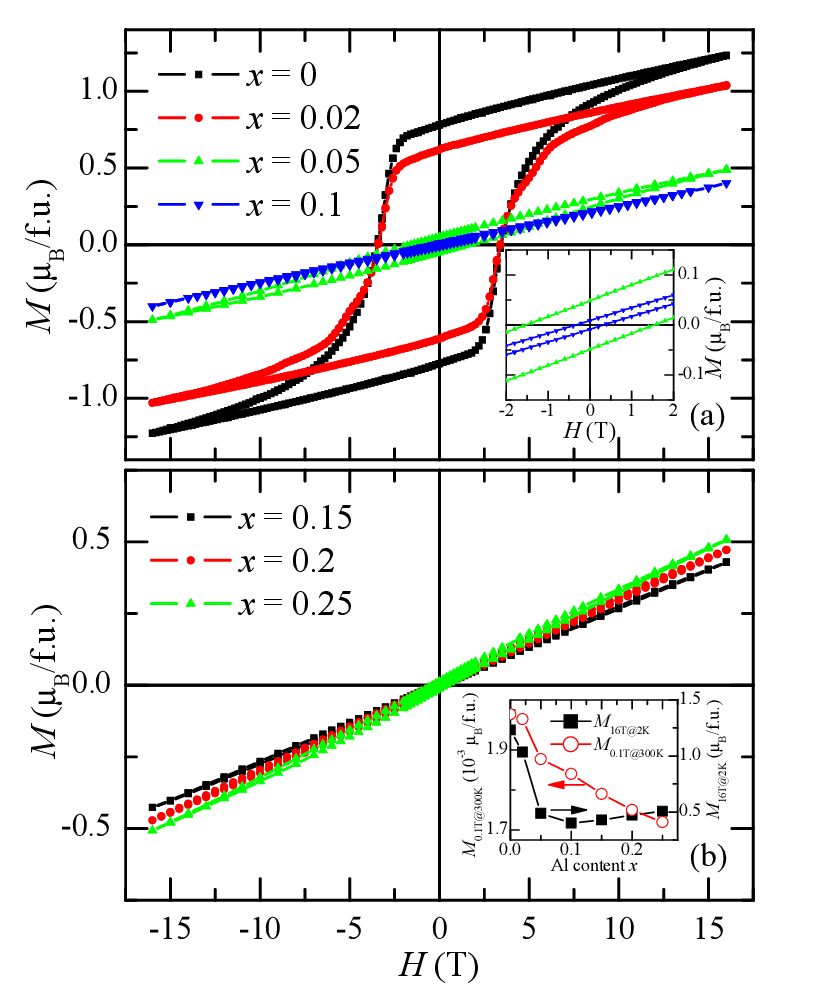}}
\caption{(Color online)The magnetization as function of the magnetic field for CaBa(Co$_{1-x}$Al$_x$)$_4$O$_7$ with 0 $\leq x \leq$ 0.25 measured at 2 K.}\label{fig:MH}
\end{figure}

The temperature dependence of the magnetization $M$($T$) for CaBa(Co$_{1-x}$Al$_x$)$_4$O$_7$ (0 $\leq x \leq$ 0.25) are shown in Fig. \ref{fig:MT}. They are measured under 0.1 T during field cooling sequence (FCC), during warming after field cooling sequence (FCW) and during warming after zero field cooling sequence (ZFC), respectively. For CaBaCo$_4$O$_7$, the magnetization shows a rapid increase upon cooling. This fact, along with the rectangle isothermal magnetization loop at 2 K (shown in Fig. \ref{fig:MH} (a)), suggests that the system enters a magnetically ordered state. The magnetic moment, which is determined from the isothermal magnetization measured at 2 K (shown in Fig. \ref{fig:MH}), is only $\sim$ 1.233 $\mu_B/f.u.$ under 16 T. This value is relatively small compared to a FM state and is consistent with the FIM ground state. A clear thermal hysteresis is observed between FCC and FCW $M$($T$) curves, indicating the first-order nature of the PM-FIM transition. All these results are consistent with previous reports. \cite{CaBaCo4O7SSC,CaBaCo4O7SRC}

\begin{figure}[t]
\center{\includegraphics[angle=0,scale=1]{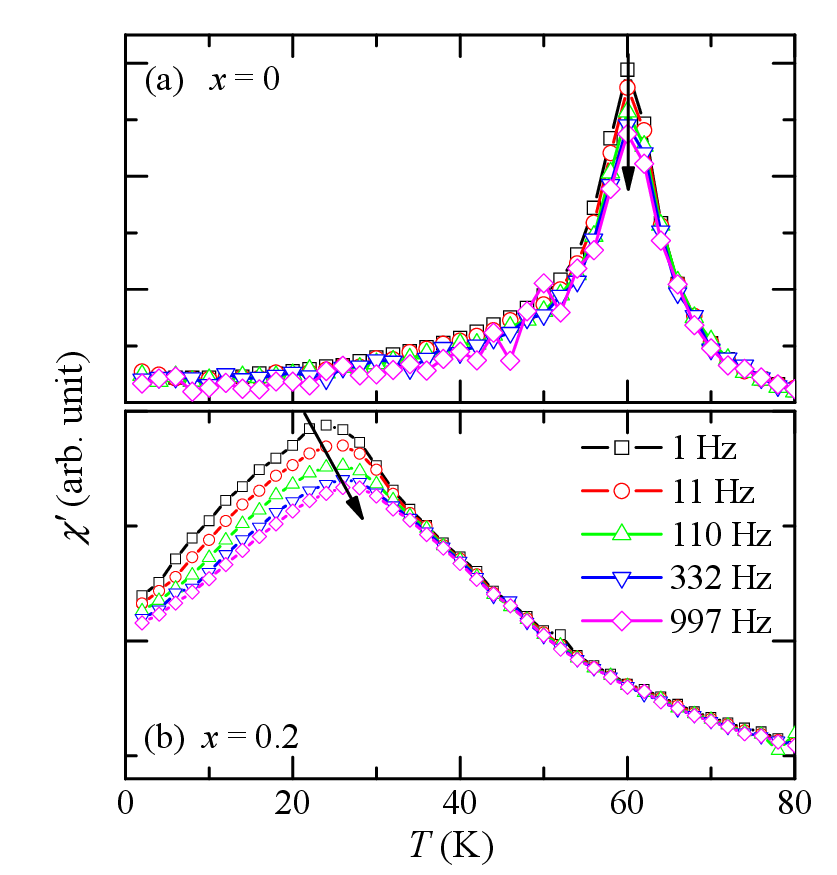}}
\caption{(Color online) The $\chi'$($T$) curves for CaBa(Co$_{1-x}$Al$_x$)$_4$O$_7$ with $x =$ 0 (a) and 0.2 (b) at different measuring frequencies.}\label{fig:AC}
\end{figure}

Upon Al substitution for Co, the transition temperature of the PM-FIM is found to be gradually suppressed toward lower temperature; it decreases from $\sim$ 60 K for $x =$ 0 to $\sim$ 30 K for $x =$ 0.1. This is accompanied with the rapid decrease of the magnetization and the coercive field (see Fig. \ref{fig:MH}), suggesting that the FIM state is gradually weakened upon Al doping. Moreover, as shown in Fig. \ref{fig:MT}, while the thermal hysteresis between FCC and FCW $M$($T$) curves is still clear visible for $x =$ 0.05, it could not be observed for $x =$ 0.1, suggesting that the PM-FIM transition changes from a first-order one to a second-order one. For $x =$ 0.05 and 0.1 the isothermal magnetization loops are no longer rectangular and the magnetization does not saturate even under an applied field of 16 T, hinting the possible existence of the spin canting in these samples.
With further increase of the Al content, the magnetic behaviors show drastic changes. For $x >$ 0.1, The $M$($H$) loops show almost linear field dependence, hinting the absence of the FIM long-range order. Meanwhile, a sharp $\lambda$ peak is observed in the FCC/FCW $M$($T$) curves, which usually means the occurrence of a spin glass state \cite{SG} or large coercivity in a long-range ordered state \cite{coercive}. 
We further performed AC susceptibility measurements on two typical samples, $x =$ 0 and 0.2, to distinguish the origin of the obvious irreversibility between DC magnetization curves (see Fig. \ref{fig:AC}). For $x =$ 0, the peak in $\chi'$($T$) keeps essentially unchanged for different measuring frequency, agreeing with the long-range FIM state. But for $x =$ 0.2, the peak in $\chi'$($T$) shifts to higher temperature with increasing frequency, characterizing the SG state. The identification of the SG state for $x >$ 0.1 is also consistent with previous observation of the SG state in CaBaCo$_3$AlO$_7$. \cite{114doping} 
All these results demonstrate that the magnetic ground states switch from the long-range FIM ordered ground state for 0 $\leq x \leq$ 0.1 to the SG state for 0.1 $< x \leq$ 0.25.

\begin{figure}[t]
\center{\includegraphics[angle=-90,scale=1]{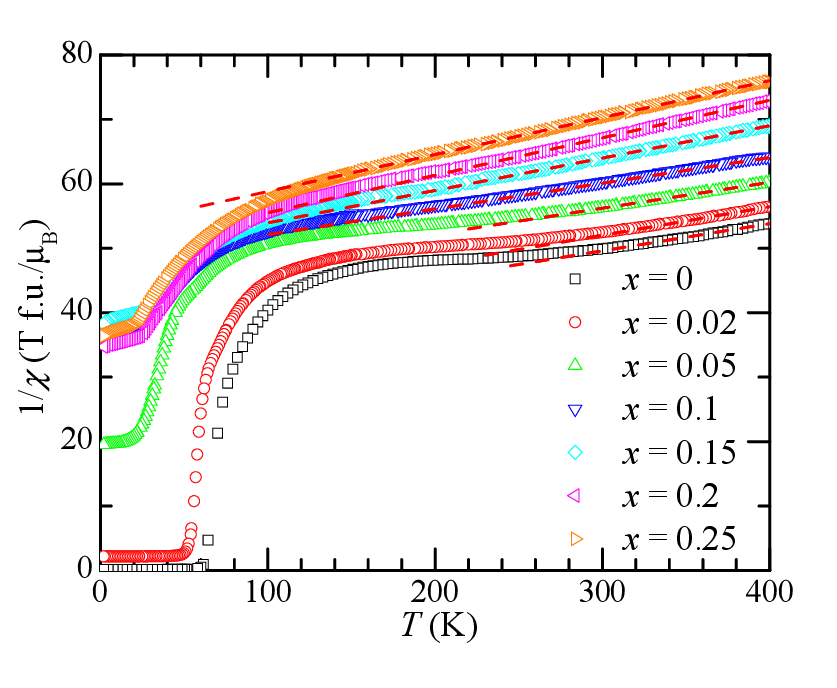}}
\caption{(Color online) The reciprocal of the susceptibility as function of temperature between 2 and 400 K under 0.1 T measured in FCC sequence for CaBa(Co$_{1-x}$Al$_x$)$_4$O$_7$ with 0 $\leq x \leq$ 0.25. The data have been vertically shifted for clarity. The dashed lines represent the Curie-Weiss fittings. The data and fitting curves are vertically shifted for clarification. }\label{fig:chi}
\end{figure}

The temperature dependence of the reciprocal susceptibility is further analyzed to investigate the evolution of the short-range magnetic correlations with Al doping.
The data are vertically shifted for clarification. As shown in Fig. \ref{fig:chi}, an upward deviation from linearity is observed in the 1/$\chi$ versus $T$ curve, suggesting the occurrence of short-range magnetic correlations. \cite{CaBaCo4O7SRC} With Al substitution for Co, the temperature corresponding to the upward deviation from linearity and the magnitude of the upward deviation gradually decrease, suggesting that Al doping gradually suppresses short-range magnetic correlations that occurs at high temperature in the parent compound. For $x =$ 0.25, the upward deviation is fully suppressed, hinting the absence of short-range magnetic correlations at high temperatures in this sample. Since the short-range magnetic correlations should be related to the geometrical frustration inherent to the system, the suppression of the short-range suggests that the geometry frustration might be partially released by Al doping.

\begin{figure}[t]
\center{\includegraphics[angle=-90,scale=1]{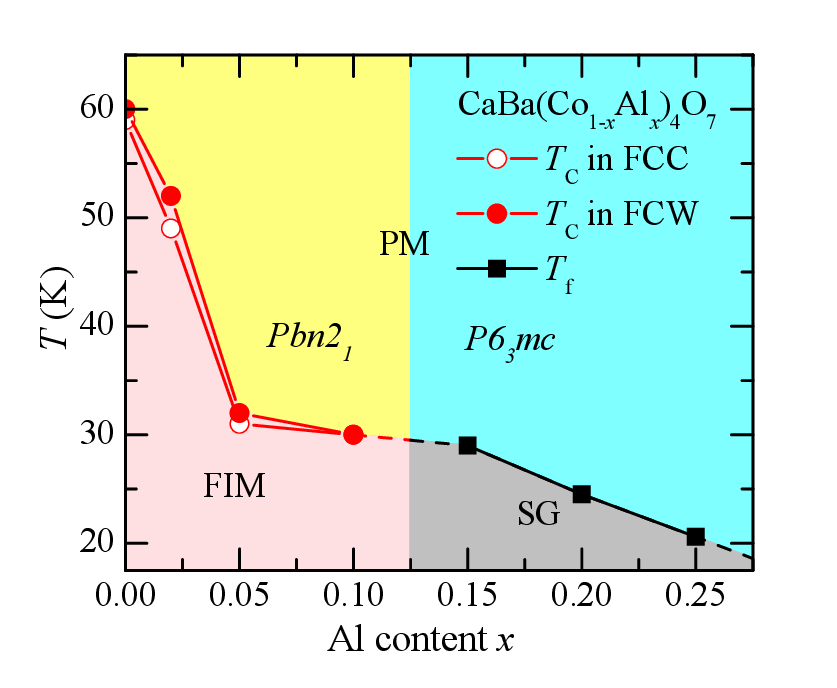}}
\caption{(Color online) The magnetic phase diagram for the CaBa(Co$_{1-x}$Al$_x$)$_4$O$_7$ with 0 $\leq x \leq$ 0.25. PM: the paramagnetic state; FIM: the ferrimagnetic state; SG: the spin-glass state. The closed circle and the open circle represent the PM-FIM transition temperature measured in FCC and FCW sequences, respectively. The filled square shows the freezing temperature of the SG state.}\label{fig:ps}
\end{figure}

We have constructed the phase diagram of the CaBa(Co$_{1-x}$Al$_x$)$_4$O$_7$ based on these observations. As shown in Fig. \ref{fig:ps}, CaBaCo$_4$O$_7$ enters a FIM ground state below $\sim$ 60 K through a first-order magnetic transition. With Al substitution for Co, the magnetic ordering temperature gradually decreases. The magnetic transition retains its first-order nature for $x \leq$ 0.05 and becomes a second-order one for $x =$ 0.1. For $x >$ 0.1, the magnetic ground state switches from a long-range ordered FIM state to a SG state. The freezing temperature of the SG state continues to decrease with further increase of the Al content.

The suppression of the ferrimagnetism and the occurrence of the SG could be understood by considering the Al doping effect. When Al is used to replace Co, it will prefer to occupy the Co$^{3+}$ sites (Co1 and Co4). While Al impurities will not directly block the FM Co chains along $b$-axis like Zn doping case, they will weaken the interaction between Co chains. Moreover, when the Co1/Co4 site is occupied by Al, the residual three Co ions could not retain their FIM configuration because of the triangular geometry constraint on these Co ions and the AFM interaction between them. \cite{CaBaCo4O7DFT}
Both effect will tend to suppress the FIM long-range order. In addition, since geometrical frustration still exist in doped samples and the chemical doping will inevitably introduce the disorder, the system eventually transits into a SG state.

It is interesting to compare the effect of Al doping to other Co site doping cases. Zn, Ga, and Al substitution for Co all induce a structural transition from orthorhombic to hexagonal symmetry. It can be seen that the structure is more susceptible against Al doping; while the orthorhombic symmetry survives up to 20\% Ga doping and 15\% Zn doping, \cite{CaBaCo4O7Ga} it becomes unstable when the Al content exceeds $x =$ 0.1. This should be ascribed to the different radius of these ions. It is known that the radius of Ga$^{3+}$ and Zn$^{2+}$ ion (6.2 nm and 7.4 nm) is closed to that of the Co$^{3+}$ ion and Co$^{2+}$ (6.1 nm and 7.45 nm) respectively while the radius of Al$^{3+}$ (5.35 nm) is smaller than that of Co$^{3+}$. \cite{lange} Compared to Al doping case, the lattice will be less affected upon Ga/Zn substitution for Co. As a result, the structure of the system is more sensitive to Al impurities.

On the magnetic properties, both Al and Ga doping donot result in the establishment of an AFM state like Zn impurity. This could be understood because Al and Ga impurities occupy the Co1/Co4 sites and leave the FM Co$^{2+}$ chain untouched. The Zn impurities also suppress the FIM order most effectively, highlighting the important role of the FM Co$^{2+}$ chains. It is interesting to note that while samples with different doping element exhibit different structure evolutions, their magnetic properties evolve with the doping level in a quite similar way. The FIM transition temperature is rapidly suppressed to $\sim$ 30 K by 5\% chemical substitution for both Al and Ga doped sample. And a generical formation of the SG state is observed when the doping level exceeds 10\% for all these doped samples. These results suggest that while the structure phase transition is accompanied with the change of the magnetic ground state in Al doped CaBaCo$_4$O$_7$ the lattice degree of freedom is relatively decoupled with the spin degree of freedom in CaBaCo$_4$O$_7$.

\section{Conclusion}

In summary, the effect of Al substitution for Co in CaBaCo$_4$O$_7$ has been studied. A structural transition from the orthorhombic symmetry to the hexagonal symmetry is observed when the Al content exceeds $x =$ 0.1. Al doping is found to suppress the FIM ground state rapidly and eventually result in a transition of the ground state into a SG state for $x >$ 0.1. Moreover, the short-range magnetic correlations, which occur at high temperatures in CaBaCo$_4$O$_7$, are found to be suppressed with increasing Al content and disappear for $x =$ 0.25. By comparing our results with other Co-site doping cases, we suggest that the lattice and the spin degrees of freedom are relatively decoupled in CaBaCo$_4$O$_7$

\section{Acknowledgments}
This work is supported by National Natural Science Foundation of China under contracts Nos. 11004198 and 11174291. Z. Q. gratefully acknowledges supports from the Youth Innovation Promotion Association, Chinese Academy of Sciences.

\newpage

\end{document}